\pdfoutput=1

\documentclass[aps,twocolumn,showpacs,floatfix,superscriptaddress]{revtex4-1}

\usepackage{dcolumn}
 \usepackage{ulem}
\usepackage{makecell}
\usepackage[utf8]{inputenc} 
\usepackage{graphicx}

\usepackage{amsfonts}
\usepackage{amsmath,bm,amssymb}

\usepackage[usenames,dvipsnames]{color}
\usepackage{kotex}

\usepackage{comment}

\usepackage{usebib} 

\begin{document}

\title{Charge density functional plus $U$ theory of LaMnO$_3$: Phase diagram, electronic structure, and magnetic interaction}


\author{Seung Woo Jang}
\affiliation{Department of Physics, Korea Advanced Institute of Science and Technology (KAIST), Daejeon 34141, Korea}

\author{Siheon Ryee} 
\affiliation{Department of Physics, Korea Advanced Institute of Science and Technology (KAIST), Daejeon 34141, Korea}

\author{Hongkee Yoon} 
\affiliation{Department of Physics, Korea Advanced Institute of Science and Technology (KAIST), Daejeon 34141, Korea}

\author{Myung Joon Han} 
\affiliation{Department of Physics, Korea Advanced Institute of Science and Technology (KAIST), Daejeon 34141, Korea}

\date{\today}

\begin{abstract}

We perform charge density functional theory plus $U$ calculation of LaMnO$_3$. While all the previous calculations were based on spin density functionals, our result and analysis show that the use of spin-unpolarized charge-only density is crucial to correctly describe the phase diagram, electronic structure and magnetic property. Using magnetic force linear response calculation, a long-standing issue is clarified regarding the second neighbor out-of-plane interaction strength. We also estimate the orbital-resolved magnetic couplings. Remarkably, the inter-orbital $e_g$-$t_{2g}$ interaction is quite significant due to the Jahn-Teller distortion and orbital ordering.

\end{abstract}


\maketitle

\section{Introduction}

LaMnO$_3$, the mother compound of colossal magneto-resistance (CMR) phenomena, is a prototypical material in which charge, spin, orbital and lattice degree of freedom are strongly coupled thereby producing a rich phase diagram \cite{salamon_physics_2001,dagotto_notitle_2003,chatterji_notitle_2004,dagotto_open_2005,tokura_critical_2006,khomskii_notitle_2014}. Bulk LaMnO$_3$ is an A-type antiferromagnetic (A-AFM) insulator with $d_{x^2}$/$d_{y^2}$-like orbital order. Its orthorhombic crystal structure has both GdFeO$_3$-type and cooperative Jahn-Teller distortions. After the celebrated observation of CMR \cite{von_helmolt_giant_1993, jin_thousandfold_1994}, LaMnO$_3$ has been a focus of tremendous research activities from various viewpoints \cite{loa_pressure-induced_2001,fuhr_pressure-induced_2008,baldini_persistence_2011,sherafati_percolative_2016, saitoh_observation_2001, ahn_effects_2000, tobe_anisotropic_2001, quijada_temperature_2001,kovaleva_spin-controlled_2004,kim_origin_2004,kovaleva_low-energy_2010,moskvin_interplay_2010,nanda_magnetic_2010,pavarini_origin_2010,flesch_orbital-order_2012}. Recently thin-film and heterostructure forms of LaMnO$_3$ have generated new excitements and possibilities \cite{smadici_electronic_2007, zhai_new_2010,garcia-barriocanal_charge_2010,garcia-barriocanal_spin_2010,gibert_exchange_2012,lee_charge_2013,di_pietro_spectral_2015,wang_imaging_2015,peng_restoring_2016,anahory_emergent_2016,chen_electron_2017} while their intriguing  behaviors driven by introducing charge carriers and/or controlling the dimensionality still need careful investigations \cite{dagotto_notitle_2003,chatterji_notitle_2004,dagotto_open_2005,tokura_critical_2006,khomskii_notitle_2014,chakhalian_colloquium_2014,bhattacharya_magnetic_2014}.

A great number of first-principles studies have been devoted to this fascinating material for the last two decades \cite{sarma_band_1995,satpathy_electronic_1996,satpathy_densityfunctional_1996,solovyev_$mathitt_2mathitg$_1996,pickett_electronic_1996,solovyev_crucial_1996,sawada_jahn-teller_1997,singh_pseudogaps_1998,sawada_orbital_1998,hu_jahn-teller_2000,su_electronic_2000,elfimov_orbital_1999,hu_jahn-teller_2000,benedetti_ab_2001,ravindran_ground-state_2002,zenia_orbital_2005,evarestov_comparative_2005, yin_orbital_2006, hashimoto_jahn-teller_2010,franchini_maximally_2012,uba_electronic_2012,lee_strong_2013,hou_intrinsic_2014,mellan_importance_2015,an_appearance_2017}. While it is certainly true that first-principles calculations contributed a lot to understanding LaMnO$_3$ and related phenomena, the fully {\it ab-initio} description is still far from  being satisfactory. Calculating the correct magnetic ground state and electronic structures have proven to be non-trivial \cite{satpathy_electronic_1996, solovyev_crucial_1996, sawada_orbital_1998, su_electronic_2000, hashimoto_jahn-teller_2010,mellan_importance_2015, kovacik_combined_2016}. For example, the microscopic origin of the A-AFM ground state and the electronic nature of its gap have been under debate. The difficulty largely arises from the technical challenges such as the determination of interaction parameters and double-counting terms when it uses DFT$+U$-type of method which has been a main workhorse in the theoretical study. Recently, a series of  investigations provided a clear understanding of the difference in between DFT$+U$ formalisms \cite{chen_density_2015,park_density_2015,chen_spin-density_2016,ryee_comparative_2017}. In this context, it is important to re-establish the $ab$-$initio$ approach for this classical material and its consequences.

In the present work, we re-examine LaMnO$_3$ within DFT$+U$ framework. In particular, we take note that all of the previous calculations have been based on the spin-density functional theory (SDFT) while the recent investigations report its unphysical nature largely coming from double countings. From a comparative study, we show that CDFT (spin un-polarized charge density functional theory)$+U$ can resolve the unphysical behavior found in SDFT$+U$. CDFT$+U$ calculation with the interaction parameters obtained from cRPA (constrained random phase approximation) is successful for describing the electronic structure and magnetic property. We also estimate the magnetic exchange coupling constant $J$ based on the CDFT$+U$ electronic structure and the response theory. It is shown that the A-AFM spin ground state is well stabilized only by nearest-neighbor interactions which is in contrast to a part of previous studies. Furthermore, our orbital-resolved $J$ calculations show that the $e_g$-$t_{2g}$ excitation channel gives rise to a significant AFM interaction which has never been clearly noticed before.

\section{Computation Details}

We performed density functional theory plus Hubbard $U$ (DFT$+U$) \cite{anisimov_band_1991,liechtenstein_density-functional_1995} calculations within GGA (generalized gradient approximation) \cite{perdew_generalized_1996}. Throughout the manuscript `SDFT' and `CDFT' is used in referring to spin and charge (spin-unpolarized) density functional scheme, respectively \cite{ryee_comparative_2017}. Thus SDFT$+U$ and CDFT$+U$ refers to the spin-polarized GGA $+U$ and spin-unpolarized GGA$+U$, respectively. For more details regarding these formulations and their comparisons, see Ref.~\onlinecite{ryee_comparative_2017}. So-called `fully localized limit (FLL)' functional form is adopted \cite{anisimov_density-functional_1993,liechtenstein_density-functional_1995}, and $U=8.0$ eV and $J_H=0.5$ eV are used for La-$4f$ states. If not mentioned otherwise, the crystal structure is fixed to the experimental structure \cite{elemans_crystallographic_1971}. All of the electronic structure and total energy calculations were carried out with {\tt `OpenMX'} software package \cite{openmx} which is based on localized pseudo-atomic orbitals (LCPAO). Vanderbilt-type norm-conserving pseudopotentials \cite{vanderbilt_soft_1990} with partial-core corrections \cite{louie_nonlinear_1982} were used to replace the deep core potentials. Three $s$, two $p$, two $d$, and one $f$ orbitals were taken as a basis set for La. Three $s$, two $p$, and one $d$ orbitals were used for Mn. Two $s$, two $p$, and one $d$ orbitals for O. $9\times9\times9$ Monkhorst-Pack $k$-point mesh was used. The geometrical optimization is further confirmed with plane-wave basis method, `{\tt VASP} (Vienna ab-initio simulation package)'.

In Sec.~III, the level splitting $\Delta$ is defined to represent the energy difference between the occupied and unoccupied orbitals. For a given orbital $\alpha$ and spin $\sigma$, the energy level of occupied states $E^{\rm{occ}}_{\alpha\sigma}$ is calculated by taking the center of mass position of projected density of states (PDOS):
\begin{align}
\begin{split}
E^{\rm{occ}}_{\alpha\sigma} = \frac{\int_{E_{f}-x}^{E_{f}}{Eg_{\alpha\sigma}(E){\rm d}E}}{\int_{E_{f}-x}^{E_{f}}{g_{\alpha\sigma}(E){\rm d}E}},\\
\end{split}
\end{align}
where $E_f$ and $g_{\alpha \sigma}(E)$ refers to the Fermi level and the calculated DOS, respectively. The minimum integration range of $E_f-x$ is chosen to include the antibonding Mn-$3d$ states only (excluding the bonding combinations). The presented results are from $x=1.8$ eV for $e_g$ and 4 eV for $t_{2g}$ states, respectively. Any of  our conclusion does not change by this choice even when the range is extended down to $-\infty$. Similarly, the unoccupied level is calculated by
\begin{align}
\begin{split}
E^{\rm{unocc}}_{\alpha\sigma} = \frac{\int_{E_{f}}^{\infty}{Eg_{\alpha\sigma}(E)dE}}{\int_{E_{f}}^{\infty}{g_{\alpha\sigma}(E)dE}},\\
\end{split}
\end{align}
and $\Delta$ is then given by $E^{\rm{unocc}}_{\alpha\sigma_{1}}-E^{\rm{occ}}_{\beta\sigma_{2}}$.

cRPA (constrained random phase approximation) \cite{aryasetiawan_frequency-dependent_2004,sasioglu_effective_2011}  calculation was performed to estimate the interaction parameters by using  `{\tt ecalj}' software package \cite{ecalj-1}. For cRPA, we used the cubic structure with lattice constant of $a=3.934$ $\rm{\AA}$ which yields the same volume with the experimental orthorhombic structure. The unphysical screening channels caused by low-lying La-$5d$ and La-$4f$ bands were removed. The $d$-$d$ screening near the Fermi energy is excluded based on so-called `$d$-$dp$ model' of MLWF (maximally localized Wannier function) technique \cite{vaugier_hubbard_2012,sakuma_first-principles_2013,amadon_screened_2014}.

The magnetic interaction, $J$, has been calculated based on magnetic force linear response theory (MFT) \cite{liechtenstein_local_1987} as extended to our non-orthogonal LCPAO method \cite{han_electronic_2004, yoon_reliability_2017}. Throughout the manuscript we used the following convention for spin Hamiltonian, 
\begin{equation} \label{Eq_Heisenberg_spin_Hamiltonian}
H=-\sum_{i\neq j} J_{ij} {\bf e}_i\cdot{\bf e}_j
\end{equation}
where ${\bf e}_{i,j}$ refers to the unit spin vectors of atomic site $i$ and $j$.

\section{Result and Discussion}
\subsection{Magnetic phase diagram}

\begin{figure}[h]
	\begin{center}
		\includegraphics[width=0.49\textwidth,angle=0]{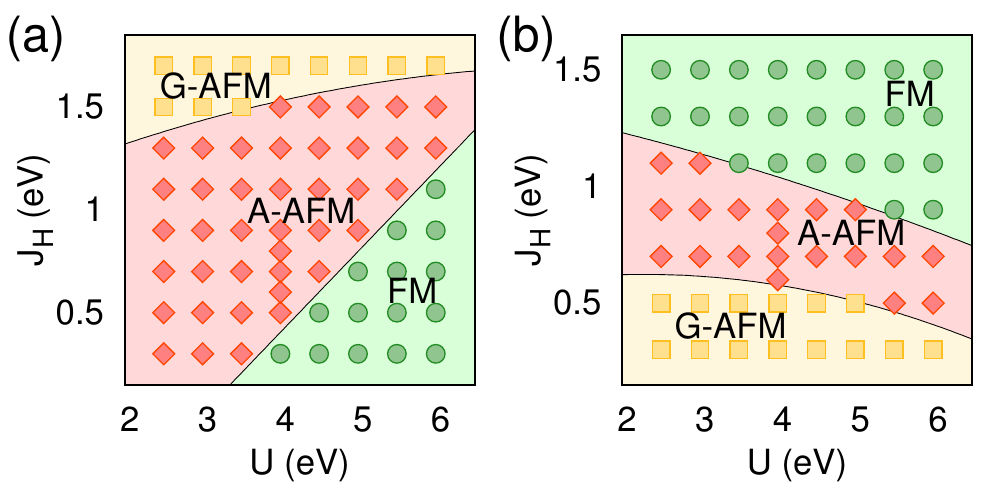}
		\caption{(Color online) The calculated magnetic phase diagram by (a) SDFT$+U$ and (b) CDFT$+U$. Yellow squares, red diamonds, and green circles show the data points corresponding to G-AFM, A-AFM, and ferromagnetic (FM) ground states spin orders, respectively.
			\label{Figure 1}}
	\end{center}
\end{figure}

Fig.~\ref{Figure 1}(a) and (b) shows the calculated magnetic phase diagram from SDFT$+U$ and CDFT+$U$, respectively. A remarkable difference is clearly noticed. In SDFT$+U$, G-AFM is stabilized in the small $U$ and large $J_H$ regime while FM is in large $U$ and small $J_H$. In CDFT$+U$, on the other hand, G-AFM is the ground state for the small $U$ and small $J_H$ values. In both phase diagrams, A-AFM order, the experimentally-known ground state, is located in between G-AFM and FM. Our cRPA estimation gives $U$=4.0 eV and $J_H$=0.7 eV, which yields the correct A-AFM ground state for both SDFT$+U$ and CDFT$+U$.

Here we note that the two widely-used standard formulations produce a quite different phase diagram for this classical material. In this regard, our result raises a serious question about the predictive power of current methodology. And in the below, we argue that CDFT$+U$ solution is physically more reliable being supported by a series of recent studies \cite{chen_density_2015,park_density_2015,chen_spin-density_2016,ryee_comparative_2017}. Notably, all of the previous DFT$+U$ calculations for bulk LaMnO$_3$ have adapted SDFT$+U$ to the best of our knowledge.

\begin{figure}[!tp]
	\begin{center}
		\includegraphics[width=0.49\textwidth,angle=0]{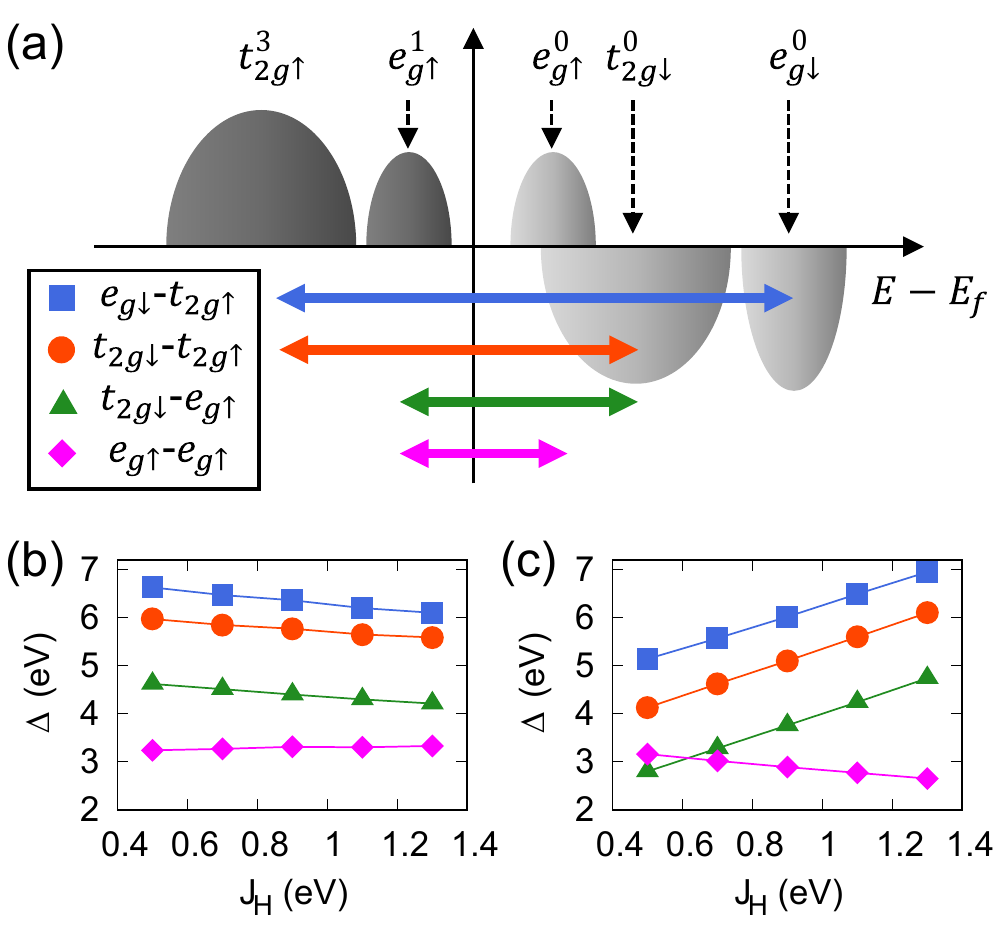}
		\caption{(Color online) (a) A schematic DOS near Fermi energy. The upper and lower panel represents the up and down spin states, respectively. The four major excitations of $e_{g\downarrow}$-$t_{2g\uparrow}$, $t_{2g\downarrow}$-$t_{2g\uparrow}$,  , $t_{2g\downarrow}$-$e_{g\uparrow}$, and $e_{g\uparrow}$-$e_{g\uparrow}$, are depicted by arrows. (b, c) The calculated $\Delta$ as a function of Hund $J_H$ by (b) SDFT$+U$ and (c) CDFT$+U$.
			\label{Figure 2}}
	\end{center}
\end{figure}

In order to understand the difference between the two formulations, we performed a systematic analysis whose results are summarized in Fig.~\ref{Figure 2}. First we define the orbital- and spin-dependent energy level splitting $\Delta$ (see Fig.~\ref{Figure 2}(a)).
For simplicity, we denote $\Delta_{e_{g\uparrow}-e_{g\uparrow}}$ by $\Delta_{\uparrow \uparrow}$, and $\Delta_{t_{2g\downarrow}-t_{2g\uparrow}}$, 
$\Delta_{e_{g\downarrow}-t_{2g\uparrow}}$, and
$\Delta_{t_{2g\downarrow}-e_{g\uparrow}}$ by $\Delta_{\downarrow \uparrow}$,
Since the magnetic interactions are approximately given by these energy differences through 
$J\sim t^2 / \Delta$ ($t$: hopping parameter), one can try to understand the phase diagrams in terms of these parameters. As clearly shown in Fig.~\ref{Figure 2}(b) and (c), the calculated $\Delta$ exhibits an opposite behavior as a function $J_H$. Namely, $\Delta_{\uparrow \uparrow}$ increases in SDFT$+U$ and decreases in CDFT$+U$. $\Delta_{\downarrow \uparrow}$ decreases in SDFT$+U$ and increases in CDFT$+U$.

This observation provides useful information to assess two different functional fomulations. It is known that $\Delta_{e_{g\uparrow}-e_{g\uparrow}}\simeq U-3J_H+\Delta_{\rm{JT}}$ and $\Delta_{t_{2g\downarrow}-t_{2g\uparrow}} \simeq  U+5J_H/2$ where $\Delta_{\rm{JT}}$ is the Jahn-Teller splitting \cite{khaliullin_orbital_2005,oles_fingerprints_2005,kovaleva_low-energy_2010}. These expressions indicate that $\Delta_{e_{g\uparrow}-e_{g\uparrow}}$ and $\Delta_{t_{2g\downarrow}-t_{2g\uparrow}}$ should be reduced 
and enlarged, respectively, as $J_H$ increases. Importantly, these features is only observed in CDFT$+U$ result of Fig.~\ref{Figure 2}(c).
Similarly, $\Delta_{e_{g\downarrow}-t_{2g\uparrow}}$ and $\Delta_{t_{2g\downarrow}-e_{g\uparrow}}$ can be expressed by $U+3J_H + \Delta_{\rm{CF}}$ \cite{excitation_energy_eg_t2g}
and $U+3J_H-\Delta_{\rm{CF}}$ \cite{excitation_energy_t2g_eg} where $\Delta_{\rm{CF}}$ is the crystal field splitting.
Both are expected to be enlarged as $J_H$ increases, which is in good agreement with CDFT$+U$ result.

The calculated $\Delta$ provides further understanding of magnetic transitions. Charge excitations in between $t_{2g\downarrow}$ and $t_{2g\uparrow}$ lead to the AFM interaction \cite{dagotto_notitle_2003,chatterji_notitle_2004, solovyev_crucial_1996, ishihara_effective_1997, oles_fingerprints_2005, solovyev_long-range_2009}, and the  enlarged $\Delta_{t_{2g\downarrow}-t_{2g\uparrow}}$ reduces the AFM coupling strength through $J\sim t^2 / \Delta$. 
Similarly, the reduced $\Delta_{e_{g\uparrow}-e_{g\uparrow}}$ enhances the FM interaction. Traditionally these two are believed to be the main magnetic interactions in LaMnO$_3$, and therefore our results in Fig.~\ref{Figure 2}(c) is consistent with the AFM-to-FM transition  in Fig.~\ref{Figure 1}(b) as a function of $J_H$. 
The transition from G-AFM to A-AFM order is related to the cooperative Jahn-Teller distortion and the orbital order. Due to the $d_{x^2}$/$d_{y^2}$-like orbital order in the $xy$ plane, the $e_g$-$e_g$ hopping, responsible for FM order, is stronger within $xy$ plane than along $z$ direction.

Another interesting feature found by comparing two phase diagrams in Fig.~\ref{Figure 1} is the calculated magnetic moment at Mn site. Although the size of moment changes is not significant, SDFT$+U$ and CDFT$+U$ exhibit an opposite trend. With a fixed $U$ value of 4.0 eV for example, the Mn moment is reduced from 3.77 to 3.68 $\mu_{B}$ as $J_H$ increases from 0.3 to 0.9 eV in SDFT$+U$. In CDFT$+U$, on the other hand, it is gradually increased from 3.30 to 3.69 $\mu_{B}$. This behavior is consistent with the opposite trend of $\Delta_{\downarrow\uparrow}$ shown in Fig.\ref{Figure 2}(b) and (c).

\begin{figure}[t]
	\begin{center}
		\includegraphics[width=0.48\textwidth,angle=0]{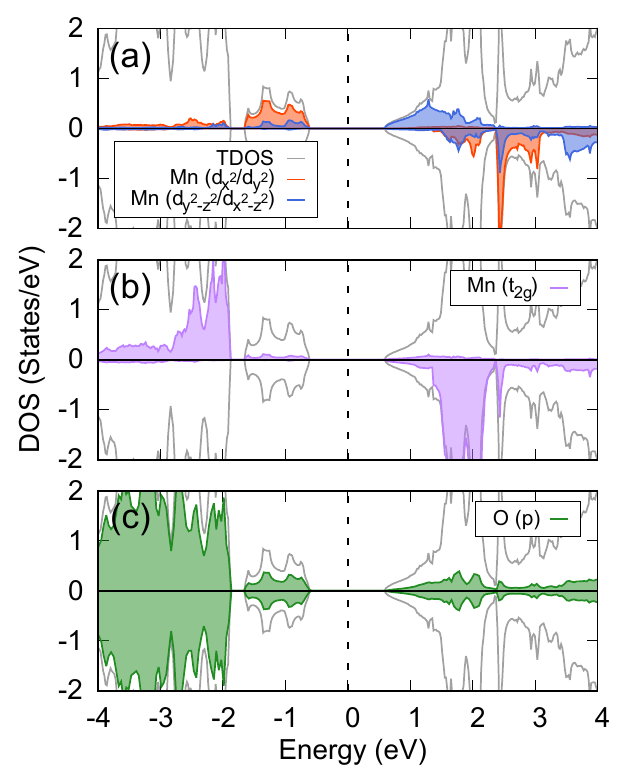}
		\caption{(Color online) (a--c) The calculated DOS corresponding to the A-AFM ground state spin order. cRPA values of $U$=4.0 and $J_H$=0.7 eV were used. The gray, red, blue, violet, and green colors represent the total DOS (divided by 4 for presentation), Mn-$d_{x^2/y^2}$, Mn-$d_{x^2-z^2/y^2-z^2}$, Mn-$t_{2g}$, and O-$2p$ states, respectively. The zero energy corresponds to the Fermi level (vertical dashed line).
			\label{Figure 3}}
	\end{center}
\end{figure}

\subsection{Electronic structure and lattice optimization}

The calculated electronic structure is presented in Fig.~\ref{Figure 3}. We once again emphasize that this is the first band structure report for LaMnO$_3$ calculated by CDFT$+U$ since all previous calculations were performed within SDFT$+U$ formalism  \cite{satpathy_electronic_1996,satpathy_densityfunctional_1996,solovyev_$mathitt_2mathitg$_1996,sawada_jahn-teller_1997,sawada_orbital_1998,elfimov_orbital_1999,hu_jahn-teller_2000,benedetti_ab_2001,ravindran_ground-state_2002,medvedeva_orbital_2002,trimarchi_structural_2005,yin_orbital_2006,ederer_structural_2007,hashimoto_jahn-teller_2010,franchini_maximally_2012,uba_electronic_2012,lee_strong_2013,hou_intrinsic_2014,mellan_importance_2015,kovacik_combined_2016}. The magnetic moment $\mu_{\textrm {Mn}}$=3.62$\mu_B$ and the gap $\Delta_{\textrm{gap}}$=1.1 eV are in good agreement with experimental values; $\mu^{\textrm{exp}}_{\textrm {Mn}}$=$3.7\pm0.1\mu_B$  \cite{elemans_crystallographic_1971}, $\Delta^{\textrm{exp}}_{\textrm{gap}}$=1.1 eV \cite{arima_variation_1993} and 1.7 eV \cite{saitoh_electronic_1995}. 
The C-type $d_{x^2}$/$d_{y^2}$ orbital order with $(\pi,\pi,0)$ ordering vector is also well reproduced. The lowest excitation is of $d$-$d$ character which supports the recent theoretical and experimental studies \cite{kovaleva_spin-controlled_2004,kovaleva_low-energy_2010,moskvin_interplay_2010}. If we use the significantly large $U\sim$ 8.0--10.1 eV obtained from constrained LDA (cLDA) \cite{satpathy_electronic_1996,solovyev_$mathitt_2mathitg$_1996}, the gap becomes a charge-transfer type as reported in a previous SDFT$+U$ study \cite{satpathy_electronic_1996}.

\begin{figure}[t]
	\includegraphics[width=0.48\textwidth,angle=0]{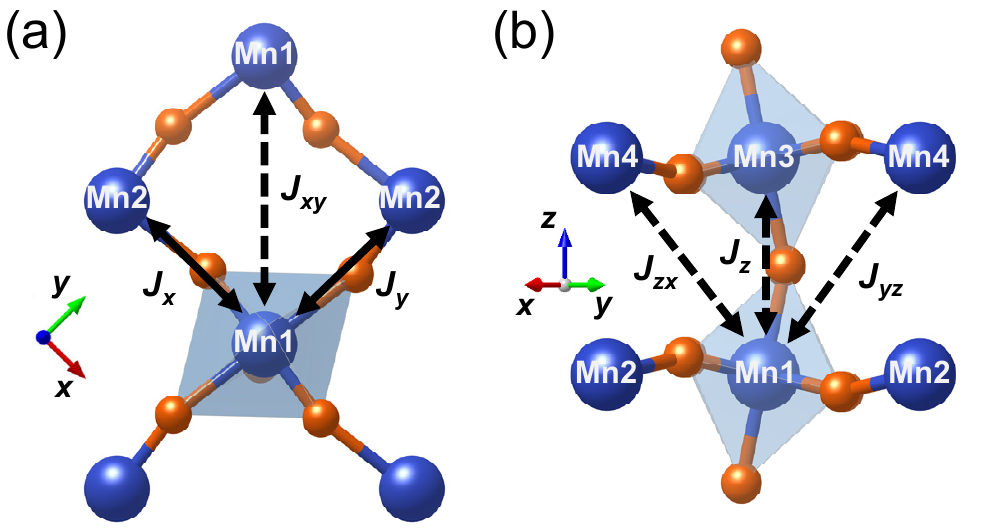}
	\caption{(Color online) (a, b) The magnetic exchange couplings in LaMnO$_3$. The in-plane and out-of-plane first-neighbor interactions are denoted by $J_{x,y}$ and $J_z$, respectively. The second neighbor couplings are denoted by $J_{xy}$ (in-plane), $J_{yz}$ (in-plane), and $J_{zx}$ (out-of-plane).}
	\label{Figure 4}
\end{figure}

Even if CDFT$+U$ provides a quite reasonable description of electronic property, it is still challenging to describe LaMnO$_3$ within the fully first-principles framework. Note that in the above the experimental lattice structure has been used. We found that the lattice optimized within CDFT$+U$ along with cRPA parameters overestimate the lattice constants and the volume by about 1--5$\%$ in comparison to experiments. The Mn-O-Mn bond angle and the orthorhombicity \cite{lee_strong_2013} is underestimated and overestimated, respectively. The similar amount of differences are also found in SDFT$+U$ result with cRPA parameters. The previous studies have been struggling with this same issue
\cite{solovyev_$mathitt_2mathitg$_1996,sawada_jahn-teller_1997,sawada_orbital_1998,hashimoto_jahn-teller_2010,mellan_importance_2015}. Due to the strong spin-charge-orbital and lattice couplings in this material, the fully self-consistent calculation for both structural and electronic property becomes quite challenging. We presume that the difficulty is largely attributed to the cRPA process which is separated from other part of the self-consistent calculation and conducted at a given lattice geometry.

\subsection{Magnetic interactions}

\subsubsection{The role of second-neighbor interactions}

Magnetic interactions in LaMnO$_3$ have long been an issue of debate particularly for the second-neighbor interaction strength.
In Ref.~\onlinecite{su_electronic_2000,evarestov_comparative_2005,solovyev_long-range_2009, franchini_maximally_2012,lee_strong_2013,hou_intrinsic_2014,mellan_importance_2015} it is argued that only the nearest-neighbor interactions ({\it i.e.,} $J_{x}$, $J_y$ and $J_z$ in Fig.~\ref{Figure 4}(a)) are important to stabilize the A-AFM order. The second neighbor $J_{zx}$ and $J_{yz}$ (see Fig.\ref{Figure 4}(a)) were either neglected or found to be small. In Ref.~\onlinecite{solovyev_$mathitt_2mathitg$_1996,solovyev_crucial_1996,kovacik_combined_2016}, on the other hand, these second neighbor interactions were claimed to play the key role in stabilizing the ground state spin order. Here we note that many different computation approaches and their combinations have been considered previously. The MFT calculation, based on LSDA (local spin-density approximation; $U$=0) spin density, and the total energy calculations, based on hybrid functional, support the significance of second-neighbor interactions \cite{solovyev_$mathitt_2mathitg$_1996,solovyev_crucial_1996,kovacik_combined_2016}. On the contrary, the total energy calculations based on {\it ab-initio} Hartree-Fock approximation reports the negligible contribution from $J_{zx}$ and $J_{yz}$ \cite{su_electronic_2000}. It is noted that each of these techniques can give a different answer for the electronic structure. For example, {\it ab-initio} Hartree-Fock produces the charge-transfer type band gap while LSDA does the Mott-Hubbard type \cite{solovyev_$mathitt_2mathitg$_1996,solovyev_crucial_1996}. With hybrid functional, a reasonable size of Mott-Hubbard gap is reproduced \cite{kovacik_combined_2016}. In this case, however, they reported that a different conclusion can be reached depending on the choices of metastable spin orders \cite{kovacik_combined_2016}.

Here we performed MFT calculation based on our CDFT$+U$ electronic structure calculated by cRPA parameters. Our result clearly shows that the second-neighbor interactions, $J_{zx}$ and $J_{yz}$, are not essential for stabilizing the A-AFM spin order. The in-plane first neighbor interaction is FM, $J_{x}=J_{y}=1.33$ meV, while the out-of-plane is AFM, $J_{z}=-0.94$ meV. The calculated $J_{xz}=J_{yz}=-0.27$ meV is AFM corresponding to 20\% and 29\% of $J_x$ and $J_z$, respectively. It is important to note that even without these interactions ($J_{zx}=J_{yz}=0$) the same ground state spin order is stabilized. The in-plane second-neighbor interaction is small enough; $J_{xy}=-0.04$ (along the short-distance $a$-axis direction) and $-0.17$ meV (along the long-distance $b$-axis direction).

\subsubsection{The orbital-decomposed results}

The magnetic orders and their phase transitions in manganites have been studied from the point of view of the competition between FM $e_g$-$e_g$ and AFM $t_{2g}$-$t_{2g}$ interactions \cite{dagotto_notitle_2003,chatterji_notitle_2004,solovyev_crucial_1996,feiner_electronic_1999,yunoki_ferromagnetic_2000,hotta_unveiling_2003,oles_fingerprints_2005,lin_theoretical_2008}. For instance, in the undoped LaMnO$_3$, a strong FM $e_{g}$-$e_{g}$ coupling wins over the in-plane ($xy$ plane) AFM $t_{2g}$-$t_{2g}$ coupling while it is weaker along $z$ direction largely due to the orbital order. This interaction profile provides a reasonable picture for the A-AFM spin ground state. From our calculation of orbitally-decomposed magnetic interactions \cite{yoon_reliability_2017}, the $J_{e_g{\textrm -}e_g}$ is indeed found to be FM; $J_{e_g{\textrm -}e_g}$=4.66 and 2.32 meV within the $xy$ plane and the out-of-plane, respectively (see Table~\ref{Table 1}). $J_{t_{2g}{\textrm -}t_{2g}}$ is AFM; $-1.56$ and $-3.03$ meV for the in- and out-of-plane, respectively. Our result is therefore consistent with the prevailing current understanding.

It is remarkable to see the significant AFM $J_{e_{g}{\textrm -}t_{2g}}$ couplings in the sense that this inter-orbital interaction has largely been ignored in the previous studies \cite{dagotto_notitle_2003,chatterji_notitle_2004,millis_cooperative_1996, ishihara_effective_1997, ahn_effects_2000, yunoki_ferromagnetic_2000, popovic_cooperative_2000, hotta_unveiling_2003, kovaleva_spin-controlled_2004, oles_fingerprints_2005, khaliullin_orbital_2005,kovaleva_low-energy_2010,sherafati_percolative_2016,snamina_spin-orbital_2018}. Our calculation shows that the in-plane $e_g$-$t_{2g}$ interaction $J_{x,y}$($e_g$-$t_{2g}$)=$-1.76$ meV is larger than $J_{x,y}$($t_{2g}$-$t_{2g}$)=$-1.56$ meV and $J_z$($e_g$-$t_{2g}$)=$-0.24$ meV. While the possibility of non-negligible $e_g$-$t_{2g}$ charge excitation was speculated in some literature \cite{solovyev_$mathitt_2mathitg$_1996, solovyev_long-range_2009,kovaleva_low-energy_2010}, our calculation provides a strong and quantitative evidence for that.

$J_{e_g{\textrm -}t_{2g}}$ interaction is not directionally symmetric. For example, while $J_{x}$ of Mn1($t_{2g}$)-Mn2($e_g$) is $-1.77$ meV, $J_{x}$ of Mn1($e_g$)-Mn2($t_{2g}$) is negligibly small (see Table \ref{Table 1}).  Due to the significant GdFeO$_3$-type distortion and orbital order, the hopping integrals between $e_g$ and $t_{2g}$ orbitals can be non-zero and the two Mn sites are no longer equivalent. This is clearly different from the case of CaMnO$_3$. As a $t_{2g}^3$ system, CaMnO$_3$ has no orbital order while it shares the GdFeO$_3$-type distortion with LaMnO$_3$. As a result,
the $e_g$-$t_{2g}$ interaction is negligible \cite{keshavarz_exchange_2017}. Our calculation shows that 
the neglect of $e_g$-$t_{2g}$ coupling in LaMnO$_3$ can be an oversimplification \cite{ishihara_effective_1997,lin_theoretical_2008}.

Our new finding of significant $e_g$--$t_{2g}$ AFM coupling has significant implication for understanding the rich phase diagram and their transitions. It is also important for the study of manganite surfaces, interfaces and thin films where the different coordination and crystal field can significantly change the magnetic interaction profiles \cite{wang_imaging_2015,anahory_emergent_2016,peng_restoring_2016,chen_electron_2017}. In the sense that the inter-orbital  couplings can even be manipulated by strain for example \cite{ahn_effects_2000,lee_strong_2013,loa_pressure-induced_2001,fuhr_pressure-induced_2008,baldini_persistence_2011,sherafati_percolative_2016}, it can have an implication for applications.

\begin{table}[t]  
	\renewcommand{\arraystretch}{1.3}
	\begin{tabular}{ c  c | c c | c c | c c}
		\hline \hline
		&\ \ & \multicolumn{2}{c|}{\ \ Mn2 : $J_{x}$ } & \multicolumn{2}{c|}{\ \ Mn2 : $J_{y}$ } & \multicolumn{2}{c}{\ \ Mn3 : $J_{z}$ }  \\
		&\ \    &\ \ $e_{g}$ &\ \ $t_{2g}$ 	&\ \ $e_{g}$ &\ \ $t_{2g}$ &\ \ $e_{g}$ &\ \ $t_{2g}$  \\
		\hline 	
		Mn1 &\ \ $e_{g}$	&\ \ 4.66  &\ \ 0.01	&\ \ 4.66  &\ \ $-$1.77  &\ \ 2.32 &\ \ $-$0.12	\\
		&\ \ $t_{2g}$ &\ \ $-$1.77  &\ \ $-$1.56 &\ \ 0.01  &\ \ $-$1.56  &\ \ $-$0.12  &\ \ $-$3.03	 \\
		\hline \hline
		
	\end{tabular}
	\caption{The calculation results of orbital-decomposed nearest-neighbor $J$. For the definition of $J_{x}$, $J_{y}$ and $J_{z}$, see Fig.\ref{Figure 3}. The unit is meV.}
	\label{Table 1}
\end{table}

\section{Summary}
We revisit a classical CMR material, LaMnO$_3$, within DFT$+U$ method. While all the previous calculations were based on SDFT$+U$, the current study clearly shows that the use of charge-only density is crucial to properly describe the electronic structure and magnetic property. It is found that the nearest-neighbor interactions are enough to stabilize the A-AFM spin ground state contrary to a part of previous studies. The  orbital-resolved $J$ calculation shows that $e_g$-$t_{2g}$ interaction is quite significant.

\section{Acknowledgments}
This work was supported by Basic Science Research Program through the National Research Foundation of Korea (NRF) funded by the Ministry of Education(2017R1D1A1B03032082).

\bibliography{References_tot}

\begin{thebibliography}{101}
\bibitem [{exc(tion)}]{excitation_energy_eg_t2g}%
  \BibitemOpen
  \href@noop {} {} (\bibinfo {year} {{For simplicity, we represented all the
  inter-orbital exchange interactions by a single parameter $J_H$. Then the
  $e_{g\downarrow}$-$t_{2g\uparrow}$ excitation,
  $(t_{2g\uparrow}^3e_{g\uparrow}^1)(t_{2g\downarrow}^3e_{g\downarrow}^1)$
  $\rightleftarrows$
  $(t_{2g\uparrow}^2e_{g\uparrow}^1)(t_{2g\downarrow}^3e_{g\downarrow}^1e_{g\uparrow}^1)$,
  can be expressed by the interaction energies of
  $t_{2g\uparrow}^3e_{g\uparrow}^1$, $t_{2g\uparrow}^2e_{g\uparrow}^1$, and
  $t_{2g\downarrow}^3e_{g\downarrow}^1e_{g\uparrow}^1$ which are given by $6U'-
  6J_H$, $3U'- 3J_H$ and $U + 9U'- 6J_H$, respectively. Thus the excitation
  energy is $U+3J_H+\Delta_{\textrm{CF}}$. Here $U'$ is on-site inter-orbital
  direct Coulomb interaction.}})\BibitemShut {NoStop}%
\bibitem [{exc( 6JH)}]{excitation_energy_t2g_eg}%
  \BibitemOpen
  \href@noop {} {} (\bibinfo {year} {{Similarly, the excitation energy of
  $(t_{2g\uparrow}^3e_{g\uparrow}^1)(t_{2g\downarrow}^3e_{g\downarrow}^1)$
  $\rightleftarrows$
  $(t_{2g\uparrow}^3e_{g\uparrow}^0)(t_{2g\downarrow}^3t_{2g\uparrow}^1e_{g\downarrow}^1)$
  is given by $6U'- 6J_H$, $3U'- 3J_H$ and $U + 9U'- 6J_H$.}})\BibitemShut
  {NoStop}%
\end{thebibliography}

\end{document}